\documentclass[prl,twocolumn,amsmath,amssymb,superscriptaddress,showpacs]{revtex4-1}
\usepackage{bm}
\usepackage{times}
\usepackage{graphicx}
\usepackage[ colorlinks,citecolor=blue,linkcolor=blue] {hyperref}

\begin{document}

\title{A quantum solution to Gibbs Paradox with few particles}

\author{H. Dong }

\affiliation{Department of Chemistry, University of California, Berkeley, CA 94720}

\author{C.Y. Cai}

\affiliation{Institute of Theoretical Physics, Chinese Academy of Sciences, Beijing,
100190, P.R. China}

\author{C.P. Sun}

\affiliation{Institute of Theoretical Physics, Chinese Academy of Sciences, Beijing,
100190, P.R. China}
\begin{abstract}
We present a fully quantum solution to the Gibbs paradox (GP) with
an illustration based on a gedanken experiment with two particles
trapped in an infinite potential well. The well is divided into two
cells by a solid wall, which could be removed for mixing the particles.
For the initial thermal state with correct two-particle wavefunction
according to their quantum statistics, the exact calculations shows
the entropy changes are the same for boson, fermion and non-identical
particles. With the observation that the initial unmixed state of
identical particles in the conventional presentations actually is
not of a thermal equilibrium, our analysis reveals the quantum origin
of the paradox, and confirm the E. J. Jaynes' observation that entropy
increase in Gibbs mixing is only due to the including more observables
measuring the entropy. To further show up the subtle role of the quantum
mechanism in the GP, we study the different finite size effect on
the entropy change and shows the works performed in the mixing process
are different for various types of particle. 
\end{abstract}


\maketitle

\textit{Introduction}-- More than one century ago, J.Willard Gibbs
\cite{Gibbs} pointed out that, if the entropy were not extensive
due to the neglection of indistinguishability of particles, there
would be an entropy increase after two ideal gases from two containers
at thermal equilibrium are mixed. On the contrary, the inverse process
would bring the system back to its original state and cause the entropy
decrease for the closed system. This apparently violates the Second
Law of Thermodynamics(SLT). However, for distinguisable particles,
this inverse process would not bring the system to the original
inital state, and not violate the SLT.

In many textbooks, e.g., the Ref.\cite{books}, the GP is claimed
to be solved by considering the indistinguishability of particle in
the view of quantum theory, and adding correction to the expression
of entropy. This mixing process will bring no entropy change for identical
particles, while additional increase for non-identical ones. However,
this pedagogic explanation includes neither the internal properties
of particle nor a quantum peculiarity of the problem (e.g., non-commutativity).
On this aspect, we still do not understand how the entropy changes
disappear only when two gases are the same, even though they can be
arbitrary similar. In the classical thermodynamic, it is always able
to assumed the substances empirically distinguished, however without
defining how small their difference is. Thus, it is expected the perfect
resolution of the GP should be referred to the quantum mechanics with
intrinsic consideration of indistinguishability, even with internal
variables of particles. This point was considered by some authors
\cite{internal2,Nieuwenhuizen06} by introducing the internal state
phenomenologically, and the discontinues of entropy change is avoided
by utilizing another thermodynamic quantity- work.

Another resolution was based on the recognition that the knowledge
of difference in particle types serves as information, which enables
extracting more work to compensate the entropy increase \cite{Jaynes1992}.
However how the internal state involve in the mixing process is still
unknown. On this sense, a fully consistent quantum description of
the GP, to our best knowledge, is still lack, especially on the indistinguishability
of the particles based on the standard quantum theory, such as the
second quantization approach with symmetrizing wave functions. Fortunately,
the quantum thermodynamics (QT) \cite{kieu2006,quan,htquan2007} has
enlighten solutions to some paradoxes in thermodynamics \cite{htquan2006,hdong2011},
e,g, the Maxwell's demon (MD) paradox \cite{MD,hdong2011,cai2011}
with quantum Szilard engine. The QT may only concern the few-particle
system, which sometimes is efficient to reveal the underlying principles
and intrinsic mechanism.

In this letter, we give a fully quantum solution to the GP involving
the symmetrization of wave function of identical particles. We consider
the mixing process of two particles confined in a infinite square
potential well, which is divided into two sub-cells by a solid wall.
Using the exact expression of the density matrix, we demonstrate explicitly
that the initial un-mixing systems in the conventional presentation
of GP are not in thermal equilibrium with respect to the whole well.
For the identical particles, we follow the standard approach of second
quantization and correctly write down the thermal equilibrium state
of an un-mixing whole system by including internal variables and symmetrizing
wave-function. This doing surprisingly restores the original version
of GP for both non- and identical particles (Boltzman particles, and
bosons/fermions) and solves the GP by showing the same entropy changes.
The detailed comparisons with the case ignoring internal state is
made to explicitly explain the origin of the paradox. For the cases
with more particles, we also demonstrate almost the same result except
for a little difference between boson and fermion, due to the finite
size effect. 

\begin{figure}[ptb]
\includegraphics[width=7cm]{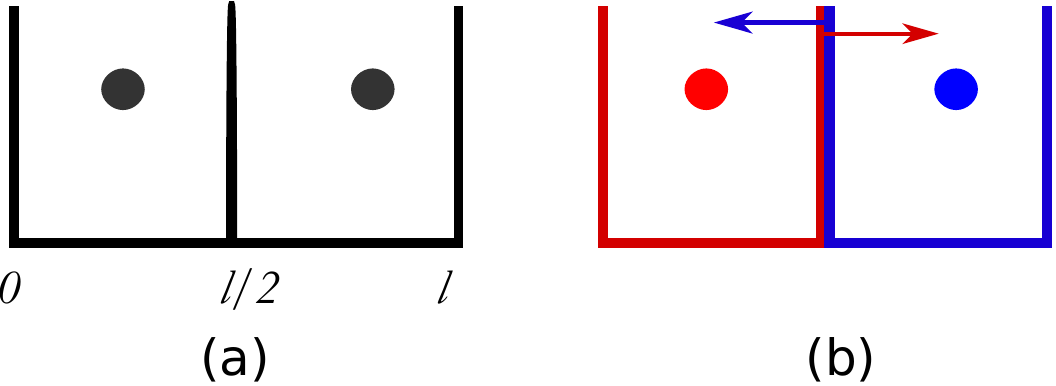} \caption{(\textit{Color Online}) Gibbs Paradox modeling with two particles
trapped in a one-dimensional square potential well of infinite depth
with width $l$. (a)The ordinary potential. The box is separated by
a central $\delta$-potential; (b) Different potentials trap different
colored particles where the colors characterizing their internal states.
The mixing is realized by isothermally moving the barrier of each
potential.}

\label{fig:gibbs_standard} 
\end{figure}

\textit{Gibbs paradox with infinite potential well-} In thermal equilibrium,
the the most probable state of many particles in a containers with
two sub-cells of equal volume usually is the situation with each cell
containing the equal number of particles. Thus, to show up the essence
of the GP, we firstly consider the two particle case, similar to the
single molecule case \cite{szilard1929} for the Szilard heat engine
in resolving the MD paradox.

Let us model the whole container as an infinite high square potential
well $V(x,l)$ ($=0$ for $x\in(0,l)$ and $\infty$ otherwhere) to
trap two particles (see Fig.\ref{fig:gibbs_standard}(a) ). The well
is initially separated by the infinite high $\delta$-potential, namely
$v\delta\left(x-l/2\right)$ with $v\rightarrow\infty$. With the
single particle eigen-wavefuctions $\left\vert \varphi_{n}^{L}\right\rangle $
( $\left\vert \varphi_{n}^{R}\right\rangle $) on left (right) cell
corresponding eigen-energies $E_{n}^{L}\left(E_{n}^{R}\right)=2\hbar^{2}n^{2}\pi^{2}/Ml^{2}$,
the eigen-wavefuctions of two distinguishable particle system read
as $\left\vert \Phi_{nm}^{uv}\right\rangle =\left\vert \varphi_{n}^{u}\right\rangle _{1}\otimes\left\vert \varphi_{m}^{v}\right\rangle _{2}$
corresponding to eigen-energies $E_{nm}^{uv}=E_{n}^{u}+E_{m}^{v}$,
for $u,v=R,L.$ With the explicit expressions of eigen-wavefuctions
given in the Supplementary Materials (SM), $\varphi_{n}^{L}(x)=$
$\langle x\left\vert \varphi_{n}^{L}\right\rangle $ and $\varphi_{n}^{R}(x)$
 are supported over the whole domain $(-\infty,\infty)$ including
$(0,l)$. This observation seems trivial, however can serves as the
key point to resolve the GP as well as the MD paradox \cite{hdong2011,cai2011}
.

To show this, we write down the thermal equilibrium state 
\begin{equation}
\rho_{\mathrm{T}}=\sum_{n,m}\sum_{u,v}\frac{1}{Z_{\mathrm{T}}}e^{-\beta E_{nm}^{uv}}\left\vert \Phi_{nm}^{uv}\right\rangle \left\langle \Phi_{nm}^{uv}\right\vert ,\label{eq:boltsman}
\end{equation}
 of two Boltzman particles with inverse temperature $\beta$ and partition
function $Z_{\mathrm{T}}$. It includes not only the terms with $\left\vert \Phi_{nm}^{RL}\right\rangle $
and $\left\vert \Phi_{nm}^{LR}\right\rangle $ for two particles in
the different sub-cells shown in Fig. \ref{fig:gibbs_standard}(a),
but also the terms with $\left\vert \Phi_{nm}^{RR}\right\rangle $
and $\left\vert \Phi_{nm}^{LL}\right\rangle $ for two particles in
the same sub-cells. If the later terms were ignored as in the conventional
description of GP, the initial un-mixing state used here would not
be in thermal equilibrium, and thus could not correctly address the
very problem in GP correctly. For non-identical particles, we can
actually assert that one specified atom in the right sub-cell and
another in the left. As emphasized by Jaynes, the specified assertion
to distinguishable particles actually reduces the system's entropy.
Such difference between non- and identical particles is the very root
of the GP. Based on this observation, we solve the GP by introducing
internal state to the identical particles to restore the original
GP in the next section.

\textit{Quantum description }-- In order to resolve the GP, we would
make the situation with equal number of particles in the both sub-cells
be the thermal state by referring certain prior information. It is
possible for non-identical particles since they could be subject to
different potentials. With the similar method, we can prepare the
initial thermal state for identical particles with internal variable.
Using the ``colors'' to label the internal
degenerate states, the blue $\left\vert b\right\rangle $ and the
red ones $\left\vert r\right\rangle $. Let the blue and red particles
be trapped respectively in the left and right sub-cells with the Hamiltonian
\begin{equation}
H=p^{2}/2M+\left\vert b\right\rangle \left\langle b\right\vert V(x,l/2)+\left\vert r\right\rangle \left\langle r\right\vert V(x-l/2,l/2),
\end{equation}
where the potential is shown in Fig.1(b). And we assume that the internal
state of the two identical particle are different. Obviously, this
assumption assigns one bit information, which brings the identical
particle with the same initial un-mixing state as non-identical one.
Therefore, the Gibbs mixing could causes entropy increase for both
non- and identical particles.

The single particle eigen-wavefunctions of the present system are
$\left\vert b\varphi_{n}^{L}\right\rangle \equiv\left\vert b\right\rangle \otimes\left\vert \varphi_{n}^{L}\right\rangle $
and $\left\vert r\varphi_{n}^{R}\right\rangle \equiv\left\vert r\right\rangle \otimes\left\vert \varphi_{n}^{R}\right\rangle $
with corresponding eigen-energy $E_{n}^{L}$ and $E_{n}^{R}$. Associating
with their quantum statistics properties, the two-particle wavefunctions
before the Gibbs mixing are symmetrized or anti-symmetrized as 
\begin{equation}
\left\vert \Psi_{nm}^{U-B\left(F\right)}\right\rangle =\frac{1}{\sqrt{2}}(\left\vert b\varphi_{n}^{L}\right\rangle _{1}\left\vert r\varphi_{m}^{R}\right\rangle _{2}\pm\left\vert r\varphi_{m}^{R}\right\rangle _{1}\left\vert b\varphi_{n}^{L}\right\rangle _{2})
\end{equation}
with the same eigen-energy $E_{nm}^{U-B\left(F\right)}=E_{n}^{L}+E_{m}^{R}$.
Here, the sign plus/minus corresponds the boson/fermion case. These
eigen-functions determine the density matrix of this system $\rho_{U-B\left(F\right)}$
(for the explicit form see the SM), similar to Eq.(1) with the partition
function $Z_{U-B\left(F\right)}=\sum_{n,m}\exp\left(-\beta E_{nm}^{U-B\left(F\right)}\right)$
.

The mixing process is carried out by moving the right side of left
potential to the position $x=l$ and the left side of the right potential
to the position $x=0$ isothermally, illustrated in Fig.1(b). After
the mixing process, two particles are constrained in the larger domain
$\left(0<x<l\right)$. The single-particle wavefunctions of this system
then become $\left\vert \phi_{n}^{s}\right\rangle \equiv\left\vert s,\phi_{n}(x,l)\right\rangle $
(degenerate for $s=b,r$,) with the same eigen-energy $E_{n}=n^{2}\pi^{2}\hbar^{2}/2Ml^{2}$.
The two-particle wavefunctions after mixing $\left\vert \Psi_{nm}^{M-B\left(F\right)}\right\rangle \sim\left\vert \phi_{n}^{b}\right\rangle _{1}\left\vert \phi_{m}^{r}\right\rangle _{2}\pm\left\vert \phi_{n}^{b}\right\rangle _{2}\left\vert \phi_{m}^{r}\right\rangle _{1}$
corresponds to eigne-energy $E_{nm}^{M-B\left(F\right)}=E_{n}+E_{m}$,
which gives the density matrix $\rho_{M-B\left(F\right)}$ with partition
function $Z_{M-B\left(F\right)}$. The entropy change in the mixing
process is simply calculated by using von Neuman entropy as $\Delta S_{B\left(F\right)}=\mathrm{Tr}\left[\rho_{U-B\left(F\right)}\ln\rho_{M-B\left(F\right)}\right]-\mathrm{Tr}\left[\rho_{M-B\left(F\right)}\ln\rho_{M-B\left(F\right)}\right]$.

We notice that the calculations about entropy change concern only
the eigen-energies, rather than the concrete form of the wavefunction.
However, the symmetry type of the wavefucntion determines the counts
of states to the partition function. For two non-identical particles
with un-symmetrized eigen-wavefunctions $\left\vert \Psi_{nm}^{U-N}\right\rangle =\left\vert b\varphi_{n}^{L}\right\rangle _{1}\otimes\left\vert r\varphi_{m}^{R}\right\rangle _{2}$,
the corresponding eigen- energy simply is $E_{nm}^{U-N}=E_{n}^{L}+E_{m}^{R}.$
Then, using the wavefunctions after mixing $\left\vert \Psi_{nm}^{M-N}\right\rangle =\left\vert b\phi_{n}\right\rangle _{1}\otimes\left\vert r\phi_{m}\right\rangle _{2}$
with the eigen-energies $E_{nm}^{M-N}=E_{n}+E_{m}$, we obtain the
same expression of the entropy increase $\Delta S_{N}$. Therefore,
it is concluded that the entropy changes in the process of mixing
two bosonic, Fermionic and non-identical particle are the same, namely
\begin{equation}
\Delta S_{B}=\Delta S_{F}=\Delta S_{N}.
\end{equation}
On this sense, GP is resolved.

\begin{figure}[ptb]
\includegraphics[width=8cm]{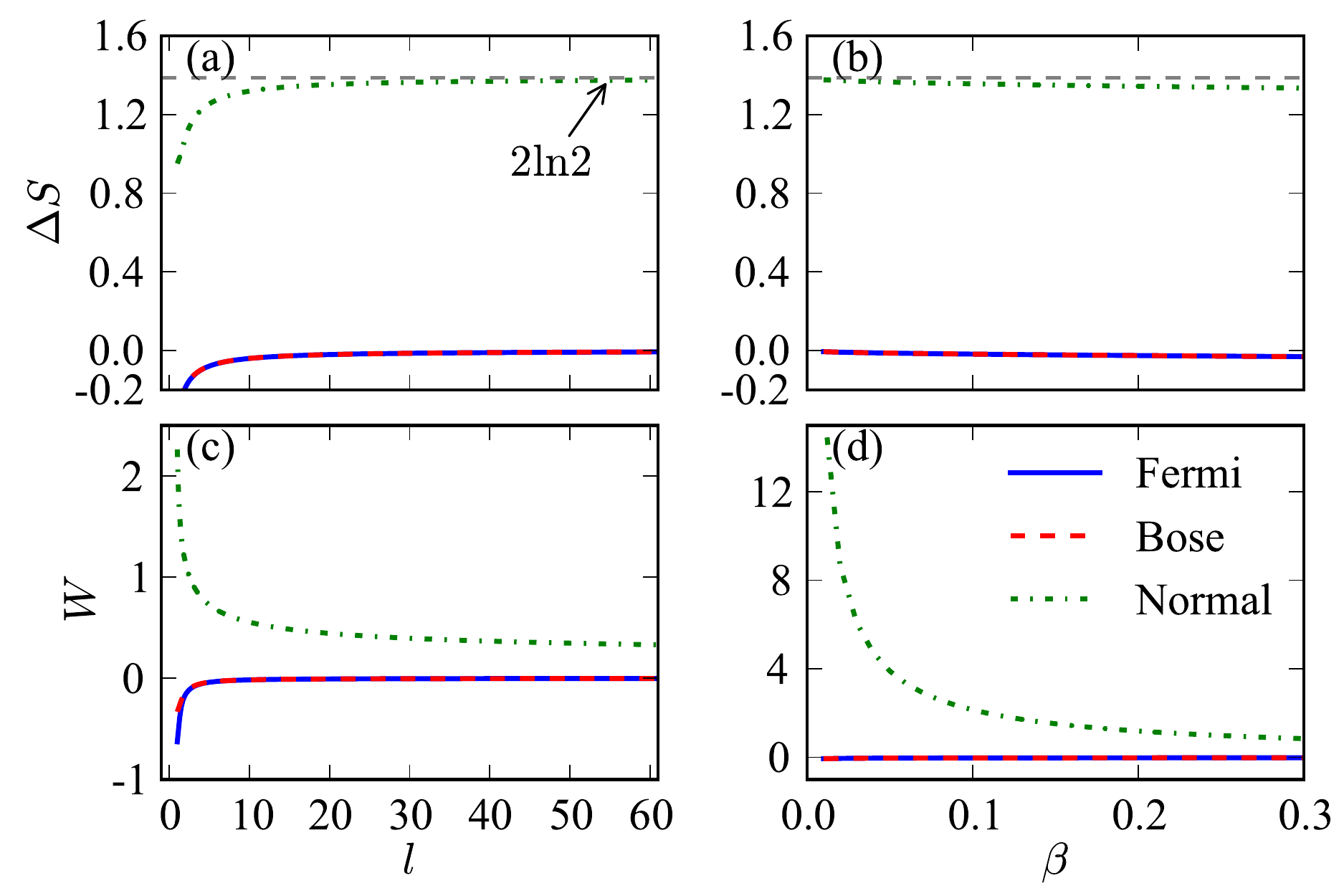} \caption{(\textit{Color online}) Entropy change in the process of mixing two
atoms as a function of (a) trap size $l$ with different inverse temperature
$\beta=0.5$ and (b) inverse temperature $\beta$ with different $l=10$
for different species: Fermi (Blue-solid), Bose(Red-dashed) and Non-identical(Green-dashdotted);
Work done as function of (c) trap size and inverse temperature(d).
The classical case of entropy change $\Delta S_{\mathrm{classical}}$
is also marked as gray-dashed line in subfigure (a) and (b).}

\label{fig:compare} 
\end{figure}

It is also meaningful to compare the present results with that without
considering the internal state for Bosonic and Fermionic particles.
In our consideration, the initial un-mixing state is a thermal state
with all the three situations in Eq.(\ref{eq:boltsman}). The mixing
process is completed by removing the central potential isothermally.
The entropy change $\Delta S'_{X}$ and work $W'_{X}=k_{B}T\left[\ln Z'_{U-X}-\ln Z'_{M-X}\right]$
are calculated by using the partition functions of un-mixing and mixed
particles $Z'_{U-X}$ and $Z'_{M-X}$ with $X=B,F,N$ . Here, we deal
with the usual setup in the conversional presentation of GP for non-identical
atoms, which is the same as that with internal state. We refer the
detailed analysis for different species to Supplement Materials. One
prominent feature is that both entropy changes and work done for different
species are different, while these are the same in the process above
with the correct considerations about internal state. We show these
differences in Fig. \ref{fig:compare}. In Fig. \ref{fig:compare}(a-b),
the entropy changes of identical particle approaches zero, while that
of non-identical particles is $2\ln2$. The discrepancy just recovers
the discontinuousness of entropy changes in conventional presentation
of GP. The similar difference in work extraction is illustrated in
Fig.\ref{fig:compare}(c-d). In the present model, no internal freedom
is probed and no prior information is added for identical particles.
In comparison with the previous presentation of GP with internal freedom,
we can conclude that the discontinuousness is caused by the different
start-point between identical and non-identical particles. 

To understand the above result, we recall the conventional presentation
of GP that the initial state of the identical particles is formally
different from the non-identical ones, which results in no entropy
change for mixing identical particles. To our understanding, the same
entropy changes are restored, solely by restoring the GP problem with
the same start points for the different types of particles. With this
observation, we conclude that the GP is rooted in the initial difference
of un-mixing state rather than the character of indistinguishability
of particles. It is also noticeable that we only utilize the quantum
definition of entropy, the von Neuman entropy, other than any phenomenological
presentations. 

\textit{Mixing entropy and work of two particle systems }-- As for
the quantum effect in our approach for GP, the finite size of the
well induces many interesting phenomenon. To explicitly demonstrate
the effect, we take the bosonic case for example. In terms of the
Theta function $\theta_{3}\left(0,q\right)=1+2\sum_{n}q^{n^{2}}$,
the partition functions before and after the Gibbs mixing are expressed\cite{derivation}
as $Z_{U-B}=\left[\theta_{3}\left(0,q\right)-1\right]^{2}/4$ and
$Z_{M-B}=\left[\theta_{3}\left(0,q^{1/4}\right)-1\right]^{2}/4$ respectively
with the parameters $q=\exp\left(-2\beta\pi^{2}\hbar^{2}/ml^{2}\right)$.
The straightforward calculation explicitly gives the entropy increase
during the mixing process as

\begin{equation}
\Delta S_{B}=2\left(\beta\partial_{\beta}-1\right)\ln\frac{\theta_{3}\left(0,q\right)-1}{\theta_{3}\left(0,q^{1/4}\right)-1}.
\end{equation}
Using the duality of the Theta-fucntion $\left(-\ln q/\pi\right)^{1/2}\theta_{3}\left(0,q\right)=\theta_{3}\left(0,\exp\left(\pi^{2}/\ln q\right)\right),$we
prove that $\theta_{3}\left(0,q\right)\rightarrow1/\sqrt{-\ln q/\pi}$
in the high temperature limit $T\rightarrow\infty$ or in the classical
limit $L\rightarrow\infty$. With this observation, the classical
result is recovered as $\lim_{L\rightarrow\infty}\Delta S_{B}=2\ln2=\Delta S_{\mathrm{classical}}.$

\begin{figure}
\includegraphics[width=8.5cm]{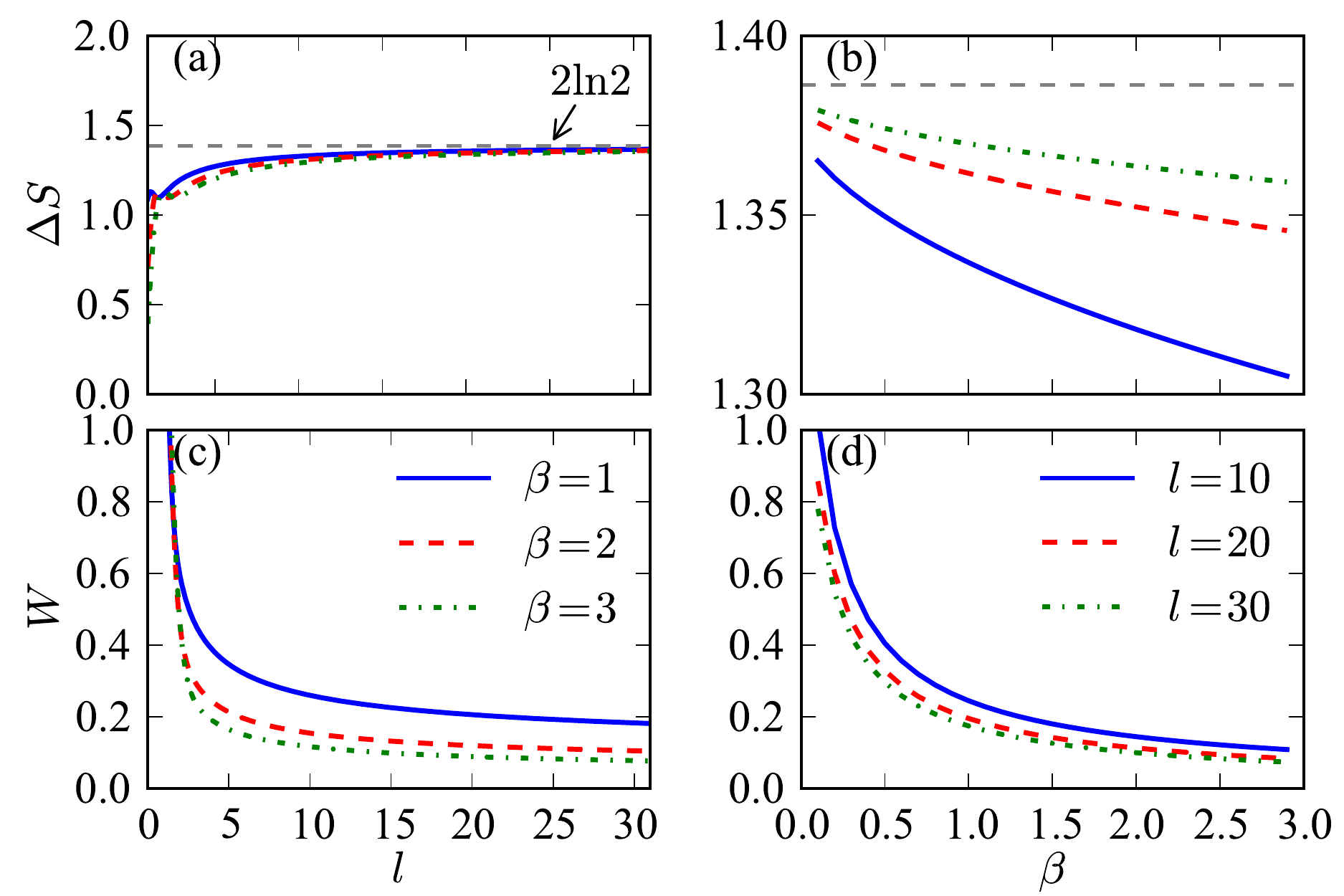} \caption{(\textit{Color online}) Entropy change in the process of mixing two
atoms as a function of (a) trap size $l$ with different inverse temperature
$\beta=1$ (blue-solid), $2$ (red-dashed) and $3$ (green-dashdotted)
and (b) inverse temperature $\beta$ with different $l=10$ (blue-solid),
$20$ (red-dashed) and $30$ (green-dashdotted). Work done as function
of (c) trap size and inverse temperature. The classical case of entropy
change is marked as gray-dashed line in subfigure (a) and (b).}

\label{fig:entropychanges} 
\end{figure}

Very quantum nature is the dependence of entropy change on size of
the trap and also on the temperature, illustrated in Fig. \ref{fig:entropychanges}(a).
The entropy change $\Delta S_{\mathrm{classical}}$ in the classical
case is marked as gray-dashed line in Fig. \ref{fig:entropychanges}(a-b).
The entropy changes tends to the classical one $\Delta S_{\mathrm{classical}}$
as the trap size approaches infinite, presented in Fig. \ref{fig:entropychanges}(a).
This confirms our theoretical analysis given above. We also show its
dependence on the inverse temperature $\beta$ in Fig. \ref{fig:entropychanges}
(b). The entropy also tends to the classical case as the inverse temperature
approaches zero$\left(\beta\rightarrow0\right)$, since the thermal
fluctuation compares well the discreteness of the energy levels in
the well. Finally, starting with the correct uses of the eigen-vectors
and -energies, we consider work done in the mixing process as the
references\cite{Nieuwenhuizen06,Jaynes1992}. The mixing process is
performed isothermally, and the work just compensates the free energy
change, namely, $W_{B}=k_{B}T\left[\ln Z_{U-B}-\ln Z_{M-B}\right]$,
which is the same for different species. This seemingly-trivial observation
also solves GP. We show in Fig. \ref{fig:entropychanges}(c-d) the
work done as function of trap size $l$ and inverse temperature $\beta$
respectively. With larger trap size, the work approaches zero. However,
it is not zero as temperature increase. Similar behavior has also
been observed in the insertion process of SHE\cite{hdong2011}. Theoretically,
the work diverges as $\sqrt{T}$ as $T\rightarrow\infty$. The detailed
derivation is presented in Supplement Materials.

\textit{Mixing entropy of 4 particle systems}-- To give prominence
to the properties of quantum statistics, we consider the mixing process
of four particles. Initially, there are two particles in each sub-cell.
The internal states of the particles in the right/left side are all
blue/red. Its partition function in thermal equilibrium is initially\cite{derivation}
\[
Z_{U-B\left(F\right)}\left(N=4\right)=\left[Z_{2}^{U-B\left(F\right)}(q)\right]^{2},
\]
where $Z_{2}^{B\left(F\right)}(q)=\frac{1}{2}[\left(Z_{1}(q^{4})\right)^{2}\pm Z_{1}(q^{8})]$
denotes the partition function of a two-particle system confined in
a square potential well with width $l/2$. For $Z_{1}(q)=\left[\theta_{3}\left(0,q\right)-1\right]/2$,
there exist two limit results: in the quantum limit with low temperature
$\beta/l^{2}\rightarrow\infty,$ we have $Z_{1}(q)\sim2q$; and in
the classical limit $\beta/l^{2}\rightarrow0$, we have $Z_{1}(q)\sim1/2\sqrt{-\ln q/\pi}-1/2$.
Together with the partition function\cite{derivation} $Z_{M-B\left(F\right)}\left(N=4\right)=\left(Z_{2}^{B\left(F\right)}(q^{1/4})\right)^{2}$
after mixing, we calculate the entropy change is 
\begin{equation}
\Delta S_{B\left(F\right)}\left(N=4\right)=2(1-\beta\partial_{\beta})\ln\frac{Z_{2}^{B\left(F\right)}(q)}{Z_{2}^{B\left(F\right)}(q^{1/4})}.
\end{equation}
In the classical limit, we have\textbf{ }$\lim_{l\rightarrow\infty}\Delta S_{B\left(F\right)}\left(N=4\right)=4\ln2$.
Indeed, when deviating from the classical limit or the lower temperature,
the entropy change obviously displays the obvious differences between
boson and fermion.

If we mix identical particles without considering the internal state,
one must start from an equilibrium state with partition function\cite{derivation}
$Z_{U-B\left(F\right)}^{^{\prime}}\left(N=4\right)=\sum_{n=0}^{N}Z_{n}^{B\left(F\right)}(q)Z_{4-n}^{B\left(F\right)}(q),$
and that $Z_{M-B\left(F\right)}^{^{\prime}}\left(N=4\right)=Z_{4}^{B\left(F\right)}(q^{1/4})$
after mixing. In the classical limit, we obtain the entropy change
as $\lim_{l\rightarrow\infty}\Delta S_{B\left(F\right)}^{^{\prime}}\left(N=4\right)=0$.
The results obtained here can be directly generated for the arbitrary
atom number $N$.

\textit{Summary}--We have resolved the Gibbs paradox in a fully quantum
framework with the correct presentations of the initial thermal states
of the working substances consisting of the particles with different
quantum statistical properties. Here, we utilize a standard quantum
description, the wave function symmetrization from the second quantization
of the particles with internal variable. The key point is our finding
that the problem in the paradox are rooted in the different uses (
somehow misuses) of the initial thermal state for identical and non-identical
particles. We use examples with two and four particles to illustrate
our comprehensive understanding for GP. We show that the entropy change
of identical and non-identical particles are the same for two particle
system, but the deference could only be found as the finite size effect
in the low temperature, or in the cases with more than two particles.

\newpage
\begin{widetext}

\part*{Supplementary Material}

In this supplemtary material, we provide the detailed calculations
of the Letter.

\ 

\textbf{\large I. Wavefunctions}

The un-mixing single particle wave-function is listed as 
\begin{equation}
\left\langle x|\varphi_{n}^{L}\right\rangle =\begin{cases}
\sqrt{\frac{4}{l}}\sin\left(2n\pi x/l\right) & 0<x<l/2\\
0 & \mathrm{otherwise}
\end{cases},
\end{equation}

\begin{equation}
\left\langle x|\varphi_{n}^{R}\right\rangle =\begin{cases}
\sqrt{\frac{4}{l}}\sin\left[2n\pi\left(x-l/2\right)/l\right] & l/2<x<l\\
0 & \mathrm{otherwise}
\end{cases}.
\end{equation}
 with the same eigen-energy $E_{n}^{L}=E_{n}^{R}=2n^{2}\pi^{2}\hbar^{2}/Ml^{2}$.
After the removing the central potential, the atoms can stay in the
whole area. The wavefunction is 
\begin{equation}
\left\langle x|\phi_{n}\right\rangle =\begin{cases}
\sqrt{\frac{2}{l}}\sin\left(n\pi x/l\right) & 0<x<l\\
0 & \mathrm{otherwise}
\end{cases},
\end{equation}
 with energy $E_{n}=n^{2}\pi^{2}\hbar^{2}/\left(2Ml^{2}\right)$.

\ 

\textbf{\large II. Mixing process of 2 atoms}

(i) \textit{With internal Freedom}

The Hamiltonian for the atoms with internal freedom is written as
\begin{equation}
H=\left|b\right\rangle \left\langle b\right|\otimes H_{b}+\left|r\right\rangle \left\langle r\right|\otimes H_{r},
\end{equation}
 where $H_{i}=p^{2}/2m+V_{i}\left(i=b,r\right)$ is the corresponding
Hamiltonian of the atom with internal state $i=b,r$. For this model,
the mixing is done by moving the right boundary of the left potential
to the right ($x=l$) and the left boundary of the right potential
to the left ($x=0$). For the Bosonic case, the partion function of
unmixing two atoms is 
\begin{eqnarray}
Z_{U-B} & = & \sum_{n,m=1}^{\infty}\exp\left[-2\beta\pi^{2}\hbar^{2}\left(n^{2}+m^{2}\right)/Ml^{2}\right]\nonumber \\
 & = & \left(\sum_{n=1}^{\infty}q^{4n^{2}}\right)\left(\sum_{m=1}^{\infty}q^{4m^{2}}\right)\nonumber \\
 & = & Z_{1}(q^{4})^{2},
\end{eqnarray}
 where $q=\exp\left(-\beta\pi^{2}\hbar^{2}/2Ml^{2}\right)$ and $Z_{1}(\tau)=\sum_{n=1}^{\infty}\tau^{n^{2}}$
is the one-particle partition function and it can be related to the
Theta function by
\[
Z_{1}(\tau)=\left[\theta_{3}\left(0,\tau\right)-1\right]/2.
\]
The partition function of mixed atoms is obtained similarly as 
\begin{equation}
Z_{M-B}=Z_{1}(q)^{2}.
\end{equation}
 The results for Fermionic and non-identical atoms is the same as
that of Bosonic case. It is clear that $q\in(0,1)$ and approaches
$1$ as $T\rightarrow\infty$ or $l\rightarrow\infty$. The Theta-fucntion
has a duality feature 
\begin{equation}
\sqrt{-\left(\ln q\right)/\pi}\,\theta_{3}\left(0,q\right)=\theta_{3}\left(0,\exp\left(\pi^{2}/\ln q\right)\right).
\end{equation}
 With this equation, we explicitely get 
\begin{eqnarray}
\theta_{3}\left(0,q\right) & = & \left(-\frac{\ln q}{\pi}\right)^{-1/2}\theta_{3}\left(0,\exp\left(\frac{\pi^{2}}{\ln q}\right)\right)\nonumber \\
 & = & \left(-\frac{\ln q}{\pi}\right)^{-1/2}\left[1+2\sum_{n=1}^{\infty}\exp\left(\frac{\pi^{2}n^{2}}{\ln q}\right)\right].
\end{eqnarray}
 The factor $\exp\left(\pi^{2}/\ln q\right)$ is no more than $1$
when $T$ or $l$ is sufficiently large, which leads to
\begin{equation}
1\leqslant1+\sum_{n=1}^{\infty}\exp\left(\frac{\pi^{2}n^{2}}{\ln q}\right)\leqslant1+\sum_{n=1}^{\infty}\exp\left(\frac{\pi^{2}n}{\ln q}\right)=\frac{1}{1-e^{\pi^{2}/\ln q}}
\end{equation}
Thus, $\sum_{n=1}^{\infty}\exp\left(\frac{\pi^{2}n}{\ln q}\right)$
approaches $0$ as $T\rightarrow\infty$ or $l\rightarrow\infty$
and the aympotic behavior of $\theta_{3}\left(0,q\right)$ as $T\rightarrow\infty$
or $l\rightarrow\infty$ is 
\begin{equation}
\theta_{3}\left(0,q\right)=\left(-\frac{\ln q}{\pi}\right)^{-1/2}\left[1+o(1)\right].
\end{equation}
 Using the above relationship, the classical limit can be obtained
as 
\begin{eqnarray}
\lim_{l\rightarrow\infty}\Delta S_{B} & = & \lim_{L\rightarrow\infty}\left(\beta\frac{\partial}{\partial\beta}-1\right)\ln\frac{\left(\sqrt{\frac{Ml^{2}}{2\beta\pi\hbar^{2}}}-1\right)^{2}}{\left(\sqrt{\frac{2Ml^{2}}{\beta\pi\hbar^{2}}}-1\right)^{2}}\nonumber \\
 & = & 2\ln2.
\end{eqnarray}

(ii) \textit{Without internal freedom }

In this section, we will discuss the mixing process of two identical
atoms without any internal freedom. In this suituation, the two atoms
has the probability to stay in the same side of the chamber, illustrated
in Fig. 1(e-f).

Bosonic Case: The eigen-functions before mixing are $\left|\Theta_{nm}^{U-B}\right\rangle _{RR}=\left[\left|\varphi_{n}^{R}\right\rangle _{1}\left|\varphi_{m}^{R}\right\rangle _{2}+\left|\varphi_{m}^{R}\right\rangle _{1}\left|\varphi_{n}^{R}\right\rangle _{2}\right]/\sqrt{2}$,
$\left|\Theta_{nm}^{U-B}\right\rangle _{RL}=\left[\left|\varphi_{n}^{R}\right\rangle _{1}\left|\varphi_{m}^{L}\right\rangle _{2}+\left|\varphi_{m}^{L}\right\rangle _{1}\left|\varphi_{n}^{R}\right\rangle _{2}\right]/\sqrt{2}$
and $\left|\Theta_{nm}^{U-B}\right\rangle _{LL}=\left[\left|\varphi_{n}^{L}\right\rangle _{1}\left|\varphi_{m}^{L}\right\rangle _{2}+\left|\varphi_{m}^{L}\right\rangle _{1}\left|\varphi_{n}^{L}\right\rangle _{2}\right]/\sqrt{2}$
for $n\not=m$ and when $n=m$ the eigen-functions $\left|\Theta_{nn}^{U-B}\right\rangle _{RR}=\left|\varphi_{n}^{R}\right\rangle _{1}\left|\varphi_{n}^{R}\right\rangle _{2}$
and $\left|\Theta_{nn}^{U-B}\right\rangle _{LL}=\left|\varphi_{n}^{L}\right\rangle _{1}\left|\varphi_{n}^{L}\right\rangle _{2}$.
The partition function is then derived as 
\begin{eqnarray}
Z'_{U-B} & = & \sum_{n,m=1}^{\infty}\left(2+\delta_{nm}\right)\exp\left[-2\beta\pi^{2}\hbar^{2}\left(n^{2}+m^{2}\right)/\left(Ml^{2}\right)\right]\nonumber \\
 & = & \frac{1}{2}\left[\theta_{3}\left(0,q^{4}\right)-1\right]^{2}+\frac{1}{2}\left[\theta_{3}\left(0,q^{8}\right)-1\right].
\end{eqnarray}
 After the mixing, the two-particle wavefunction are $\left|\Theta_{nm}^{M-B}\right\rangle =\left[\left|\phi_{n}\right\rangle _{1}\left|\phi_{m}\right\rangle _{2}+\left|\phi_{m}\right\rangle _{1}\left|\phi_{n}\right\rangle _{2}\right]/\sqrt{2}$
for $n\not=m$ and when $n=m$ $\left|\Theta_{nn}^{M-B}\right\rangle =\left|\phi_{n}\right\rangle _{1}\left|\phi_{n}\right\rangle _{2}$.
The partition function is 
\begin{eqnarray}
Z'_{M-B} & = & \sum_{n,m=1}^{\infty}\frac{1}{2}\left(1+\delta_{nm}\right)\exp\left[-\beta\pi^{2}\hbar^{2}\left(n^{2}+m^{2}\right)/\left(2Ml^{2}\right)\right]\nonumber \\
 & = & \frac{1}{8}\left[\theta_{3}\left(0,q\right)-1\right]^{2}+\frac{1}{4}\left[\theta_{3}\left(0,q^{2}\right)-1\right].
\end{eqnarray}

Fermionic Case: The function before mixing should be anti-symmeterized
as $\left|\Theta_{nm}^{U-F}\right\rangle _{RR}=\left[\left|\varphi_{n}^{R}\right\rangle _{1}\left|\varphi_{m}^{R}\right\rangle _{2}-\left|\varphi_{m}^{R}\right\rangle _{1}\left|\varphi_{n}^{R}\right\rangle _{2}\right]/\sqrt{2}$,
$\left|\Theta_{nm}^{U-F}\right\rangle _{LL}=\left[\left|\varphi_{n}^{L}\right\rangle _{1}\left|\varphi_{m}^{L}\right\rangle _{2}-\left|\varphi_{m}^{L}\right\rangle _{1}\left|\varphi_{n}^{L}\right\rangle _{2}\right]/\sqrt{2}$
with $n\neq m$ and $\left|\Theta_{nm}^{U-F}\right\rangle _{RL}=\left[\left|\varphi_{n}^{R}\right\rangle _{1}\left|\varphi_{m}^{L}\right\rangle _{2}-\left|\varphi_{m}^{L}\right\rangle _{1}\left|\varphi_{n}^{R}\right\rangle _{2}\right]/\sqrt{2}$.
\begin{eqnarray}
Z'_{U-F} & = & \sum_{nm}\left(2-\delta_{nm}\right)\exp\left[-2\beta\pi^{2}\hbar^{2}\left(n^{2}+m^{2}\right)/\left(Ml^{2}\right)\right]\nonumber \\
 & = & \frac{1}{2}\left[\theta_{3}\left(0,q^{4}\right)-1\right]^{2}-\frac{1}{2}\left[\theta_{3}\left(0,q^{8}\right)-1\right].
\end{eqnarray}
 After the mixing, the two-particle wavefunction are $\left|\Theta_{nm}^{M-F}\right\rangle =\left[\left|\phi_{n}\right\rangle _{1}\left|\phi_{m}\right\rangle _{2}-\left|\phi_{m}\right\rangle _{1}\left|\phi_{n}\right\rangle _{2}\right]/\sqrt{2}$
with $n\neq m$. The partition function is 
\begin{eqnarray}
Z'_{M-F} & = & \sum_{nm}\frac{1}{2}\left(1-\delta_{nm}\right)\exp\left[-\beta\pi^{2}\hbar^{2}\left(n^{2}+m^{2}\right)/\left(2mL^{2}\right)\right]\nonumber \\
 & = & \frac{1}{8}\left[\theta_{3}\left(0,q\right)-1\right]^{2}-\frac{1}{4}\left[\theta_{3}\left(0,q^{2}\right)-1\right].
\end{eqnarray}

\ 

\textbf{\large III. Mixing process of 4 atoms }

(i) \textit{With internal freedom}

At beginning, there are two blue atoms (whose internal states are
$\left|b\right\rangle $) in the left compartment and two red atoms
(whose internal states are $\left|r\right\rangle $) in the right
compartment. Thus, the initial partition function is
\begin{equation}
Z_{U-B\left(F\right)}\left(N=4\right)=\left(Z_{2}^{B\left(F\right)}(q)\right)^{2},
\end{equation}
where $Z_{2}^{B\left(F\right)}(q)$ denotes the connonical partition
function of a two-particle system confined in a square protential
well with width $l/2$. The explicit expression of $Z_{2}^{B\left(F\right)}(q)$
is
\begin{equation}
Z_{2}^{B\left(F\right)}(q)=\frac{1}{2}\left(\left(Z_{1}(q^{4})\right)^{2}\pm Z_{1}(q^{8})\right).
\end{equation}
The concrete form of the un-mixing gas is rewritten as 
\begin{align}
Z_{U-B\left(F\right)}\left(N=4\right) & =\frac{1}{4}\left(\left(Z_{1}(q^{4})\right)^{2}\pm Z_{1}(q^{8})\right)^{2}\nonumber \\
 & =\frac{1}{4}\left(\left(\frac{\theta_{3}(0,q^{4})-1}{2}\right)^{2}\pm\frac{\theta_{3}(0,q^{8})-1}{2}\right)^{2}.
\end{align}

After mixing, these four atoms are confined in a infinity high square
well with width $l$. The partition function is simply
\begin{align}
Z_{M-B\left(F\right)}\left(N=4\right) & =\left(Z_{2}^{B\left(F\right)}(q^{1/4})\right)^{2}\nonumber \\
 & =\frac{1}{4}\left(\left(Z_{1}(q)\right)^{2}\pm Z_{1}(q^{2})\right)^{2}\nonumber \\
 & =\frac{1}{4}\left(\left(\frac{\theta_{3}(0,q)-1}{2}\right)^{2}\pm\frac{\theta_{3}(0,q^{2})-1}{2}\right)^{2}.
\end{align}
Therefore the entropy change is
\begin{equation}
\Delta S_{B\left(F\right)}=\left(1-\beta\partial_{\beta}\right)\ln\frac{Z_{M-B\left(F\right)}\left(N=4\right)}{Z_{U-B\left(F\right)}\left(N=4\right)}.
\end{equation}
Using the property of Theta function, it is not difficult to show
that under the classical limit ($T\rightarrow\infty$ or $l\rightarrow\infty$
), the entropy change tends to be $4\ln2$ for both Boson and Fermion
systems.

\begin{table}
\begin{centering}
\begin{tabular}{|c|c|}
\hline 
 & $Z_{n}^{\mathrm{B\left(F\right)}}(q)$\tabularnewline
\hline 
\hline 
$n=2$ & $\frac{1}{2}\left(Z_{1}(q)^{2}\pm Z_{1}(q^{2})\right)$\tabularnewline
\hline 
$n=3$ & $\frac{1}{6}\left(Z_{1}(q)^{3}\pm3Z_{1}(q^{2})Z_{1}(q)+2Z_{1}(q^{3})\right)$\tabularnewline
\hline 
$n=4$ & $\frac{1}{24}\left(\begin{array}{c}
Z_{1}(q)^{4}+3Z_{1}(q^{2})^{2}\pm6Z_{1}(q^{4})\\
\pm6Z_{1}(q^{2})Z_{1}(q)^{2}+8Z_{1}(q^{3})Z_{1}(q)
\end{array}\right)$\tabularnewline
\hline 
\end{tabular}
\par\end{centering}

\caption{\label{Z_bose}The relationship between $Z_{n}^{\mathrm{B\left(F\right)}}(q)$
and $Z_{1}(q)$ for $n=2,3,4$. }
\end{table}

(ii)  \textit{Without internal freedom}

In this section, we discuss the mixing process which is the same as
the one in the last section except that the atoms considered here
do not contain internal freedom. In this circumstance, the initial
state is a mixing of 5 situations, i.e., the $n$-th situation is
that there are $n$ atoms in the left compartment while the others
are in the right compartment. Thus, the initial partition function
is
\begin{equation}
Z'_{U-B\left(F\right)}\left(N=4\right)=\sum_{n=0}^{4}Z_{n}^{B\left(F\right)}(q)Z_{4-n}(q),
\end{equation}
where $Z_{n}^{B\left(F\right)}(q)$ denotes the connonical partition
function of a $n$-particle system confined in a square protential
well with width $l/2$. The relationship between $Z_{n}^{B\left(F\right)}(q)$
and $Z_{1}(q)$ can be found in Tab.(\ref{Z_bose}) .

After the mixing process, the system becoms to be a well containing
four identity atoms and  the partition function is
\begin{equation}
Z'_{U-B\left(F\right)}\left(N=4\right)=Z_{4}(q^{1/4})=\frac{1}{24}\left(\begin{array}{c}
Z_{1}(q)+3Z_{1}(q^{2})^{2}\pm6Z_{1}(q^{4})\\
\pm6Z_{1}(q^{2})Z_{1}(q)^{2}+8Z_{1}(q^{3})Z_{1}(q)
\end{array}\right).
\end{equation}
 This partion function can also be related to the Theta function by
the ways mentioned above. The entropy change during this process is
\[
\Delta S_{B\left(F\right)}^{'}\left(N=4\right)=\left(1-\beta\partial_{\beta}\right)\ln\frac{Z'_{M-B\left(F\right)}\left(N=4\right)}{Z'_{U-B\left(F\right)}\left(N=4\right)},
\]
which can also be proved that $\lim_{L\rightarrow\infty}\Delta S_{B\left(F\right)}^{'}\left(N=4\right)=0$
by using the property of the Theta function. 

\end{widetext}

\end{document}